\documentclass[12pt]{article}

\newenvironment{numberedlist}
{\begin{list}{\makebox[20pt]{\hss(\arabic{itemno})\enspace}}
             {\usecounter{itemno}\labelwidth 20pt}}{\end{list}}

\newcounter{itemno}

\newcounter{itemno1}

\newcounter{itemno2}

\newcounter{exno}

\newcounter{defno}

\newenvironment{defn}{\refstepcounter{defno}\medskip \noindent {\bf
Definition \thedefno.\ }}{\medskip}

\newcommand{\sep}{\;\vert\;}

\newcommand{\oprove}{\vdash\kern-.6em\lower.7ex\hbox{$\scriptstyle O$}\,}

\newcommand{\Pscr}{{\cal P}}

\newcommand{\pderivation}{{\cal P}\kern -.1em\hbox{\rm -derivation}}
\newcommand{\pderivationl}{{\cal P}\kern -.1em\hbox{\em -derivation}}
\newcommand{\pderivable}{{\cal P}\kern -.1em\hbox{\rm -derivable}}
\newcommand{\pderivablel}{{\cal P}\kern -.1em\hbox{\em -derivable}}
\newcommand{\pderivations}{{\cal P}\kern -.1em\hbox{\rm -derivations}}
\newcommand{\pderivability}{{\cal P}\kern -.1em\hbox{\rm -derivability}}

\newcommand{\all}{\forall}
\newcommand{\some}{\exists}

\newcommand{\ie}{{\em i.e.}}

\newsavebox{\lpartfig}
\newsavebox{\rpartfig}

\newenvironment{exmple}{
 \begingroup \begin{tabbing} \hspace{2em}\= \hspace{3em}\= \hspace{3em}\=
\hspace{3em}\= \hspace{3em}\= \hspace{3em}\= \kill}{
 \end{tabbing}\endgroup}

\newcommand{\sand}{sand} % choice disjunction
\newcommand{\pand}{pand} % choice disjunction
\newcommand{\lb}{\langle}
\newcommand{\rb}{\rangle}

\newcommand{\prove}{exec} % choice conjunction
     {\\* \hspace*{\fill} \end{trivlist}}

\newcommand{\seqand}{\land^s}

\newcommand{\seqandq}[2]{\prec_{#1}^{#2}}

\renewcommand{\seqand}{\land}
\renewcommand{\seqandq}[2]{\land_{#1}^{#2}}

\begin{document}

\begin{center}
{\Large {\bf For-loops in Logic Programming}}
\\[20pt]

{\bf Keehang Kwon}\\
Dept. of Computer Eng. \\
DongA University \\
%051-200-7784 \\
 khkwon@dau.ac.kr\\

\vspace{1em} 

%{\bf Jiseung Kim}\\
%Dept. of Industrial Logistics \\ 
%KyungIl University \\
%051-200-7784 \\
%jiskim@kiu.ac.kr\\

\end{center}

%% </local definitions here>

%\begin{document}
%\maketitle

\noindent {\bf Abstract}: 
Logic programming  has  traditionally lacked devices for
 expressing  iterative tasks.
 To overcome this problem, this paper proposes  
  iterative goal formulas of the form 
 $\seqandq{x}{L} G$  where  $G$ 
is a goal, $x$ is a variable,  and $L$ is a list. $\seqandq{x}{L}$ is called a parallel bounded
quantifier. These goals allow us to specify
the following   task:  iterate $G$ with $x$ ranging over all the elements of $L$.

{\bf keywords:}  for-loop,  iteration, bounded quantifier, computability logic

\newcommand{\proloop}{{Prolog$^{forloop}$}}

%\begin{multicols}{2}

%% </local definitions here>

\section{Introduction}\label{sec:intro}
 
Logic programming has traditionally lacked mechanisms that permit some tasks
 to be  iterated. This deficiency is an outcome of
using a weak logic as the basis for logic programming. 
Lacking  looping
constructs, logic programming relies on recursion to perform  iterative 
goal tasks.
 One of the disadvantages of this approach is that  even simple iterative goal tasks are difficult to read, 
write and reason about.  Also, iteration can be directly implemented much
more efficiently than recursion. 

To deal with this deficiency,
our approach in this paper involves the direct enrichment of the underlying intuitionistic 
logic to a fragment of Computability Logic(CL) in \cite{Jap03,Jap08} to allow for  iterative  goals.
 A parallel iterative  goal  is of the form 
$\seqandq{x}{L}  G$ 
 where $G$ is a goal, $x$ is a variable,  and $L$ is a list.
Executing this goal has the following intended semantics: iterate $G$ with $x$ ranging over all 
elements of the list $L$.
 All executions must succeed for executing $\seqandq{x}{L}  G$  to succeed.

    An
illustration of this  facet is provided by the following definition of the
relation which sequentially writes all the elements in a list:

\begin{exmple}
$      write\_list(L)$ ${\rm :-}$ \> \hspace{6em}      $write("List: ") \seqand$\\ 
 \> \hspace{6em} $(\seqandq{x}{L} write(x)).$
\end{exmple}
\noindent
which replaces  the tedious logic program  shown below:

\begin{exmple}
$      write\_list(L)$ ${\rm :-}$ \> \hspace{8em}      $write("List: "), $\\ 
 \> \hspace{8em} $ write\_list1(L).$ \\
$      write\_list1([])$.\\ 
$      write\_list1([X|T])$ ${\rm :-}$ \> \hspace{8em}      $write(X),$\\ 
 \> \hspace{8em} $ write\_list1(T).$
\end{exmple}

The body of the new definition above contains  an iterative goal. 
 As a particular example, solving the query $write\_list([1,2,3])$ would result in
solving the goal $\seqandq{x}{[1,2,3]}$, after writing $List:$.  The given goal will
succeed after writing $1,2,3$ in sequence. 

     As seen from the example above,  iterative  goals can be used to perform looping tasks.
    This paper proposes  \proloop, an extension of Prolog with  iterative operators in goal formulas.

There are some previous works \cite{Apt96,Joa} that
have advocated the use of bounded quantifiers.
Although their motivation is similar to ours, the difference is that
their approach stays within the framework of Prolog.
In other words,  bounded quantifiers are just 
syntactic sugars and must be transformed to lengthy Prolog  codes before execution.

Our approach overcomes this inefficiency: bounded quantifiers are now legal
and can be implemented in a direct, efficient way, \ie, without
translation to Prolog.

In this paper we present the syntax and semantics of this extended language, 
show some examples of its use. 

The remainder of this paper is structured as follows. We describe \proloop\
based on a first-order Horn clauses with bounded quantifiers in
the next section. In Section \ref{sec:modules}, we
present some examples.
Section~\ref{sec:conc} concludes the paper.

\section{The Language}\label{sec:logic}

The language is a version of Horn clauses
 with  iterative goals. It is described
by $G$- and $D$-formulas given by the syntax rules below:
\begin{exmple}
\>$G ::=$ \>  $A \sep G \land  G \sep  \some x\ G \sep G\seqand G \sep \seqandq{x}{L} G$ \\   
\>$D ::=$ \>  $A  \sep G \supset A\ \sep \all x\ D  \sep  D \land D $\\
\end{exmple}
\noindent

In the rules above,  $x$  represents a variable, $L$ represents a list of terms, and  
$A$  represents an atomic formula.
A $D$-formula  is called a  Horn
 clause with  iterative goals. 
 
In the transition system to be considered, $G$-formulas will function as 
queries and a set of $D$-formulas will constitute  a set of 
instructions. For this 
reason, we refer to a $G$-formula as a query, to a set of $D$-formula 
as an instruction set. 

 We will  present an operational 
semantics for this language as inference rules.  To be specific, we encode such inference rules  as
theories in the (higher-order) logic of task, \ie,  a simple variant of Computability Logic \cite{Jap03}.
Below the expression $A\ sand\ B$ denotes a sequential conjunction of the task $A$ and the task $B$ and
the expression $A\ pand\ B$ denotes a parallel conjunction of the task $A$ and the task $B$.

These rules in fact depend on the top-level 
constructor in the expression,  a property known as
uniform provability\cite{Mil89jlp,MNPS91}. 

\begin{defn}\label{def:semantics}
Let $G$ be a goal and let $\Pscr$ be a finite set of  instructions.
Then the notion of   executing $\lb \Pscr,G\rb$ -- executing $G$ relative to $\Pscr$ -- 
 is defined as follows:

\begin{numberedlist}

\item $\prove(\Pscr,A)$ if $A$ is identical to an instance of a program clause in $\Pscr$.

\item  $\prove(\Pscr,A)$ if (an instance of a program clause in $\Pscr$ is of the form
 $G_1 \supset A$) $pand$ $\prove(\Pscr, G_1)$.

\item $\prove(\Pscr,G_1 \land G_2)$ if $\prove(\Pscr,G_1)$  $\pand$ 
  $\prove(\Pscr,G_2)$. Thus, the two goal tasks  must be done in parallel and both tasks 
must succeed for the current task to succeed.

\item $\prove(\Pscr,\exists x G_1)$  if (select the true term $t$)
 $\sand$ $\prove(\Pscr,[t/x]G_1)$. Typically, selecting the true term 
                      can be achieved via the unification process.

\item $\prove(\Pscr, \seqandq{x}{nil} G)$. The current execution terminates with a  success.

\item  $\prove(\Pscr,\seqandq{x}{[a_1,\ldots,a_n]} G)$ if 
$\prove(\Pscr, [a_1/x]G)$ $\pand$   $\prove(\Pscr,\seqandq{x}{[a_2,\ldots,a_n]} G)$ .

%\item  $\prove(\Pscr,\rep{cond} G)$ if $G$ $\sand$ $(
%$\neg cond$ $\cor$ $(cond$ $\sand$ $\prove(\Pscr,\rep{cond} G)))$. 

\end{numberedlist}
\end{defn}

\noindent  
In the above rules, the symbols $\seqandq{x}{L}$  provides  iterations: they allow for the repeated 
 conjunctive execution of
the instructions. We plan to investigate whether
 this semantics is sound and complete with respect to CL.

An alternative yet tedious way to giving semantics of our language is by transformation to plain logic programming.
For example, our loop construct  $\seqandq{x}{L}$ can be defined by introducing a recursive auxiliary predicate
such as $ write\_list1$ in Section 1. This method is  discussed in detail in \cite{Joa}.

\section{Examples }\label{sec:modules}

An example is provided by the 
following ``factorial'' program. 

\begin{exmple}
$fact(0,1).$ \% base case \\  
$fact(X+1,XY+Y)$ ${\rm :-}$ \> \hspace{9em} $fact(X,Y).$\\
\end{exmple}

Our language in Section 2 permits iterative
 goals.  An example of
 this construct is provided by the 
 program which 
does the following  tasks:  output 10!, 11!,  12!, 13! sequentially:

\begin{exmple}
\> $query1:$.\\
\> $\seqandq{N}{[10,11,12,13]}$ \% for i= 10 to 13 begin\\
\>$(fact(N,O) \seqand$\\
\>$write(N) \seqand write('factorial\ is:') \seqand$ \\
\>$write(O))$ \% for end\\
\end{exmple}
\noindent 
 For example, consider a goal $query1$.
Solving this goal  has the effect of executing $query1$ 
 with respect to  the factorial program for four times.

Our language in Section 2 permits variables to appear in the list
 in iterative goals.  These variables can be used only for controlling iteration and
must be instantiated at run-time. An example of
 this construct is provided by the 
 program which 
does the following iterative tasks: read a number $N$ from the
user,  and then repeatedly output the factorials of the numbers from 1 to $N$. 

\begin{exmple}
\> $query2:$.\\
\>$(read(N) \seqand$ \\
\> $\seqandq{x}{[1..N]}$ \% for $x$= 1 to N begin\\
\>$(fact(x,O) \seqand$\\
\>$write(x) \seqand write('factorial\ is:') \seqand$ \\
\>$write(O))$ \% for end\\
\end{exmple}
\noindent 
In the above, note that $[1..N]$ is a shorthand notation for $[1,2,\ldots,N]$.

%\begin{exmple}
%\> $query2:$.\\
%\>$(read(N) \seqand$ \\
%\> $\rep{N = Esc}}$ \% for $x$= 1 to N begin\\
%\>$(fact(N,O) \seqand$\\
%\>$write(N) \seqand write('factorial\ is:') \seqand$ \\
%\>$write(O))$ \% for end\\
%\end{exmple}

\section{Conclusion}\label{sec:conc}

In this paper, we have considered an extension to logic programming with  
 iterations in goals. This extension allows goals of 
the form $\seqandq{x}{L} G$  where $G$ is a goal, $x$ is a variable and $L$ is a list of terms. 
These goals are 
 particularly useful for the bounded  looping executions of instructions, making logic programming
more concise, more readable, and more friendly to imperative programmers.

    Although   iterative goals do provide a significant gain in expressive
elegance, some tasks -- with dynamic termination conditions -- cannot be expressed at all using them.  
We plan to  look at some variations \cite{Joa} such as the $fromto$ statements in the future
to improve expressibility. 

Regarding implementing our language, the handling of  bounded quantifications
does not pose any major
complications. The treatment of a goal of the form $G_1 \seqand G_2$
that is indicated by the operational semantics does not forbid
 $G_1$ and $G_2$ to be processed sequentially, as is done in most Prolog implementations.

\section{Acknowledgements}

This work  was supported by Dong-A University Research Fund.

\bibliographystyle{plain}

\begin{thebibliography}{1}
\bibitem{Apt96}
K.~Apt, ``Arrays, bounded quantification and iteration in logic and
constraint logic programming'', Science of Computer Programming, vol.26, pp.133--148, 1996.


\bibitem{Joa}
J.~Schimpf, ``Logical loops'', ICLP, pp.224--238, 2002.

\bibitem{Jap03}
G.~Japaridze, ``Introduction to computability logic'', Annals  of Pure and
 Applied  Logic, vol.123, pp.1--99, 2003.

\bibitem{Jap08}
G.~Japaridze,   ``Sequential operators in computability logic'',
 Information and Computation, vol.206, No.12, pp.1443-1475, 2008. 
 
\bibitem{Jap12a}
G.~ Japaridze, ``A new face of the branching recurrence of computability logic'',
Applied Mathematics Letters (to appear).
    
\bibitem{kh10a}
K.~Kwon and S.~Hur,   ``Adding sequential conjunctions to {P}rolog'', 
 IJCTA, vol.1, No.1, pp.1-3, 2010. 

\bibitem{Mil89jlp}
D.~Miller, ``A logical analysis of modules in logic programming'', Journal of
  Logic Programming, vol.6, pp.79--108, 1989.

\bibitem{MNPS91}
D.~Miller, G.~Nadathur, F.~Pfenning, and A.~Scedrov, ``Uniform proofs as a
  foundation for logic programming'', Annals of Pure and Applied Logic, vol.51,
  pp.125--157, 1991.

\end{thebibliography}

%\profile*{}{}% without picture of author's face

%\end{multicols}

\end{document}